# Effect of Spin-Orbit Interaction and In-Plane Magnetic Field on the Conductance of a Quasi-One-Dimensional System

Yuriy V. Pershin, James A. Nesteroff, and Vladimir Privman

*Center for Quantum Device Technology,*
*Department of Physics and Department of Electrical and Computer Engineering,*
*Clarkson University, Potsdam, New York 13699-5721, USA*

We study the effect of spin-orbit interaction and in-plane effective magnetic field on the conductance of a quasi-one-dimensional ballistic electron system. The effective magnetic field includes the externally applied field, as well as the field due to polarized nuclear spins. The interplay of the spin-orbit interaction with effective magnetic field significantly modifies the band structure, producing additional sub-band extrema and energy gaps, introducing the dependence of the sub-band energies on the field direction. We generalize the Landauer formula at finite temperatures to incorporate these special features of the dispersion relation. The obtained formula describes the conductance of a ballistic conductor with an arbitrary dispersion relation.

Recently, there have been numerous studies, both theoretical and experimental, of the properties of quasi-one-dimensional systems [1-8]. The motivation behind this interest has been the observation of conductance quantization. Most quasi-one-dimensional systems, or Quantum Wires (QW), are created by a split gate technique in a two-dimensional electron gas (2DEG) [6,7]. When a negative potential is applied to the gates, the electrons are depleted underneath. Thus, a one-dimensional channel or constriction is created between two reservoirs, in this case the 2DEG. For ballistic transport to occur [1], this constriction should be less than the electron mean free path, and have a width of the order of de Broglie wavelength [6-8]. When these conditions are satisfied, the electrons will move ballistically in the lateral direction and are confined transversely. The transverse confinement creates a discrete set of modes in the channel.

The explanation for conductance quantization is found by using a non-interacting electron model. With a small bias applied across the channel, the electrons move from one reservoir to the other. Due to the transverse confinement in channel, the electrons are distributed, according to the Fermi-Dirac distribution, among various sub-bands in the channel. The calculation of the conductance has been summarized in the Landauer–Büttiker formalism [6-8]. Each one of the sub-bands contributes $2e^2/h$ to the conductance.

The spin-orbit (SO) interaction is described by the Hamiltonian [9-12],

$$H_{SO} = \frac{\hbar}{4m^2c^2}\left(\vec{\nabla}V \times \vec{p}\right)\cdot\vec{\sigma}, \qquad (1)$$

where $\vec{\sigma}$ represents a vector of the Pauli spin matrices, $m$ is the free electron mass, $\vec{p}$ is the momentum operator, and $\vec{\nabla}V$ is the gradient of a potential, proportional to the electric field acting on the electron. When dealing with crystals, there are two main types of spin-orbit interaction. The Dresselhaus spin-orbit interaction [11] appears as a result of the asymmetry present in certain crystal lattices, e.g., zinc blende structures. The Rashba spin-orbit interaction [13] arises due to the asymmetry associated with the confinement potential and is of interest because of the ability to electrically control the strength of this interaction. The latter is utilized, for instance, in the Datta-Das spin transistor [14]. The Hamiltonian for the Rashba interaction is written [13] as

$$H_{SO} = i\alpha\left(\sigma_x \frac{\partial}{\partial y} - \sigma_y \frac{\partial}{\partial x}\right), \qquad (2)$$

where $\alpha$ is the coupling constant. In this paper, we limit our consideration to the systems with only the Rashba interaction. Incorporation of Dresselhaus interaction into our calculations is straightforward.

In the simplest case when the external magnetic field is applied in the plane of the heterostructure and the spin-orbit coupling is neglected, each sub-band in the channel is spin-split. The effect this field has on conductance is that each sub-band now contributes $e^2/h$ to the conductance, as observed, e.g., in [3]. Another possible source of spin-splitting are nuclear spins, which we incorporate into our model within an effective-field approximation. We introduce the total magnetic field as $\vec{B} = \vec{B}_{ext} + \vec{B}_N$, where $\vec{B}_{ext}$ is the external magnetic field and $\vec{B}_N$ is the field produced by the polarized nuclear spins. The maximum nuclear field in GaAs can be as high as 5.3 T, in the limit when all nuclear spins are fully polarized [15]. This high level of nuclear spin polarization has been achieved experimentally. For example, optical pumping of nuclear spins in a 2DEG has yielded nuclear spin polarization of the order of 90% [16]. A similarly high polarization, 85%, has been created by quantum-Hall edge states [17]. It is anticipated that the effect of the nuclear spin polarization on the conductance is similar to the effect of in-plane magnetic field [18]. The characteristic energy scale of the interaction between conduction electrons and polarized nuclear spins is of the order of the Fermi energy, what makes possible to confine electrons into low-dimensional electron structures using modulation of nuclear spin polarization [19]. This idea was developed in [5,20-22]. We assume that $\vec{B}_N$ is parallel to the external magnetic field $\vec{B}_{ext}$ [16].

Recent studies of 2DEG-based systems [23-25] have been focused on including the effects of the spin-orbit interaction on the conductance with perfect channel transmission, $T(E)=1$. Other studies [8,26-28] have dealt with similar effects in hybrid ferromagnetic-semiconductor systems but with transmission coefficients less than 1. Moreover, ballistic transport with spin-orbit interaction on a cylindrical surface was considered in Ref. [29]. In our calculations we assume perfect transmission through the channel and focus on the effects of the interplay of spin-orbit interaction and in-plane magnetic field on the conductance at finite temperatures. Our approach also takes into account the influence of polarized nuclear spins. We show that the spin-orbit interaction breaks the symmetry of sub-bands in certain cases and manifests itself as unique features in conductance dependence on the gate voltage.

Let us consider the energy spectrum of a quasi-one-dimensional system in which we take into account the effects of the spin-orbit interaction and applied in-plane magnetic field. The Hamiltonian inside the conductor will then be

$$H = \frac{p^2}{2m^*} + U(y) - i\alpha\sigma_y \frac{\partial}{\partial x} + \frac{g^*\mu_B}{2}\vec{\sigma}\cdot\vec{B} \quad . \quad (3)$$

Here $p$ is the electron momentum in the $x$ direction, $m^*$ is the electron effective mass, $U(y)$ is the electron confining potential in the $y$ direction, $g^*$ is the effective g-factor and $\mu_B$ is the Bohr magneton. The Rashba term, $H_R$, defined in (2), was reduced for the motion along $x$ only. We assume that the total magnetic field experienced by electrons is in-plane, $\vec{B} = B_x\hat{x} + B_y\hat{y}$. It should be emphasized that the magnetic field does not enter into (3) through the vector potential in our approximation.

We consider solutions of the Schrödinger equation inside the constriction that are separable, of the form

$$\psi = e^{ikx}\phi(y)\begin{pmatrix}\varphi_\uparrow \\ \varphi_\downarrow\end{pmatrix}, \quad (4)$$

where $\phi(y)$ is the wave function for the transverse modes (due to the confinement potential) and $\varphi_{\uparrow,\downarrow}$ are the spinor components for spin up and down, respectively. The eigenvalue problem can be solved to obtain

$$E^n_\pm = \frac{\hbar^2 k^2}{2m^*} + E^{tr}_n \mp \left[\left(g^*\mu_B B/2\right)^2 + g^*\mu_B \alpha kB\sin\theta + (\alpha k)^2\right]^{1/2} . \quad (5)$$

In this expression, the up and down spin states in the eigenbasis are denoted by $\pm$, $\theta$ is the angle of the magnetic field relative to the electron transport through the wire, $B\sin\theta$ represents the $y$-component of the total magnetic field, and $E^{tr}_n$ is the spectrum of transverse sub-bands. Assuming the parabolic confinement potential in the $y$-direction, we have $E^{tr}_n = \hbar\omega(n+1/2)$. The energy spectrum corresponding to Eq. (5) is illustrated in Fig. 1 (a)-(c) for various values of $\theta$. Recently, similar energy-spectrum calculations have been reported in [30].

It is interesting to note some of the properties of these sub-bands. In the case of $\theta$ equal to $0$, Fig. 1 (c), or $\pi/4$, Fig. 1 (b), a gap appears at $p = 0$ between two spin-split bands, which is not observed when $\theta = \pi/2$. The sub-bands in Figs. 1 (b) and (c) also contain local extrema, due to the Rashba term.

Our goal has been to calculate the overall influence of the spin-dependent interactions in the Hamiltonian on the finite temperature conductance. To do this, we make use of the Landauer-Büttiker formalism. We consider a model of a quantum wire, which consists of two electron reservoirs, with chemical potentials $\mu_L$ and $\mu_R$, separated by a conductor. This conductor is assumed to be devoid of scatterers so that the transmission coefficient, $T(E)$, is unity. If a bias, $eV = \mu_L - \mu_R$, is applied across the contacts, such that $\mu_L > \mu_R$, then the total current, I, through the conductor can be written as the difference between the current flows in the forward and reverse directions, see Ref. [7], page 52,

$$I = \frac{e}{2\pi}\sum_{n,s}\int_{-\infty}^{\infty} v_{ns}[\vartheta(v_{ns})f(E,\mu_L) + \vartheta(-v_{ns})f(E,\mu_R)]dk .\qquad(6)$$

Here $f(E(k),\mu_{L,R})$ is the Fermi-Dirac distribution function for electrons in the left and right reservoir, $v_{ns}$ is the electron velocity given by $\hbar^{-1}\partial E_s^n/\partial k$, with $s = \pm$ denoting the spin state, $E_s^n$ is given by Eq. (5), and $\vartheta(v)$ is the step function. The summation over $n$ and $s$ includes contributions from all the sub-bands in the channel.

As previously discussed, the band structure of our system exhibits a number of local extrema which have to be taken into account in order to calculate the conductance using Eq. (6). For example, if we consider a single sub-band with an arbitrary number of local extrema, as shown in Fig. 2, then the calculation of the conductance can be accomplished by splitting up the integral in Eq. (6) between each extremal point in the sub-band. This gives the relation,

$$I_{n,s} = \frac{e}{h}[\int_{\infty}^{E_0^{n,s}} f(E,\mu_R)dE + \int_{E_0^{n,s}}^{E_1^{n,s}} f(E,\mu_L)dE + \int_{E_1^{n,s}}^{E_2^{n,s}} f(E,\mu_R)dE + ...],\qquad(7)$$

where $I_{n,s}$ is the contribution from the sub-band labeled by $n$ and $s$, while $E_i^{n,s}$ is the $i^{th}$ energy extremum in that sub-band. If we assume that the applied bias is small, then we can Taylor-expand the integrals in Eq. (7) in terms of $eV$. By taking only the first-order terms, summing over different sub-bands and using the relation for conductance, $G = I/V$, we obtain the result

$$G = \frac{e^2}{h}\sum_{n,s}\sum_i \beta_i^{n,s} f(E_i^{n,s}) .\qquad(8)$$

Here, the sum is calculated over the extremal points of all the sub-bands, and $\beta_i^{n,s}$ is either +1 for a minimum or −1 for a maximum. So, we see that maximum points in a sub-band actually reduce the conductance.

Experimentally, the conductance plateaus are observed by changing the potential applied to the gates. This changing potential can be viewed as a shift of the chemical potential. The calculated values for $G$, using (8), and the corresponding temperature dependences, are shown in Fig. 3 as functions of the chemical potential, $\mu$. In Fig. 3 (a), we see that as $\mu$ increases, the zero-temperature conductance increases in steps of $e^2/h$. This is half of the increment that would be obtained without the Zeeman splitting of the energy sub-bands [8]. As the temperature is increased we observe a smearing of the conductance plateaus, since electrons coming in from the reservoirs no longer have a sharp step-like energy distribution at the chemical potential.

When the applied magnetic field has a non-vanishing x-component, new interesting effects occur in the conductance plateaus, especially at low temperatures. In Figs. 3 (b) and (c), we see the conductance plateaus for $\theta = \pi/4$ and $\theta = 0$, respectively. In the case of $\theta = \pi/4$, we can see from Fig. 1 (b) that the energy sub-bands are split, so as the chemical potential increases, the first contribution to the conductance will be the lowest minimum in Fig. 1 (b), and the second contribution occurs when the chemical potential passes the second minima of the same sub-band. As $\mu$ further increases and passes the local maximum, the conductance decreases by $e^2/h$. By this mechanism, peaks of the conductance are formed, as demonstrated in Fig. 3 (b), (c). We note that temperature-smeared curves look very similar to those in the much investigated 0.7 anomaly phenomenon [3], though additional investigations would be needed before speculating that the present mechanism could be an alternative to other explanations this effect offered in the literature.

It should be emphasized that the external magnetic field is not necessary to observe the conductance peculiarities discussed in this paper. Similar effects can be obtained due to interactions that mix spin states. The simplest example is polarized nuclear spins with a non-zero in-plane component of their spin-polarization. The influence of the nuclear spin polarization on conductance of a quasi-one-dimensional system is already included in the present formalism. However, the nuclear spin polarization is dynamic because of the spin diffusion and relaxation processes. Since the electron equilibration time scales are much shorter than the time scales of dynamics of the nuclear spin system, the adiabatic approximation can be applied [5]. Within this approximation, the effective magnetic field due to the polarized nuclear spins can be considered quasi-static and relations presented above can be used to describe the conductance.

In summary, this paper has focused on the interplay of various in-plane magnetic field components, Rashba spin–orbit interaction, and finite temperature on the conductance. We have found that the angle of the magnetic with respect to the conductance channel has a significant effect on the conductance. The variation of the angle of the field generates gaps in the energy sub-band structure, that control the pattern of the conductance variation as the chemical potentials are varied by applied gate voltages.

We gratefully acknowledge helpful discussions with S. N. Shevchenko. This research was supported by the NSF (grant DMR-0121146), and by the NSA and ARDA (under the ARO contract DAAD 19-02-1-0035).

(a)

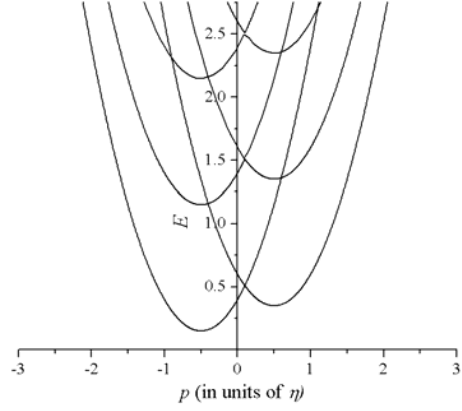

(b)

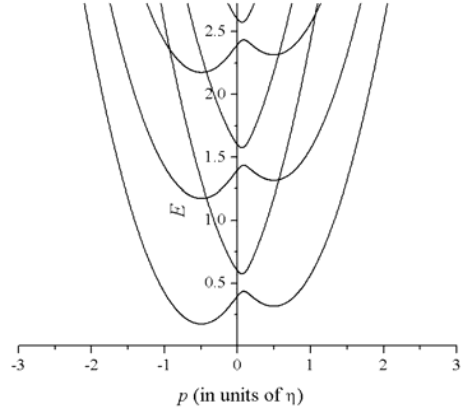

(c)

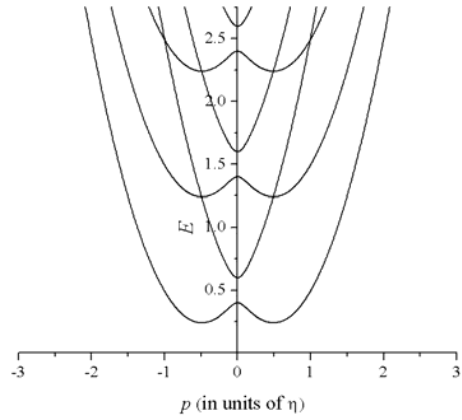

**FIG. 1.** Dispersion relations: energy, $E$, in units of $\hbar\omega$, for different values of $\theta$, (a) $\theta = \pi/2$, (b) $\theta = \pi/4$, and (c) $\theta = 0$, as a function of the momentum, $p$, where $\eta$ is defined as $\eta = (2m^*\hbar\omega)^{1/2}$. These plots were constructed using parameters: $g^*\mu_B B/(2\hbar\omega) = 0.1$ and $\alpha(2m^*/(\hbar^3\omega))^{1/2} = 1$.

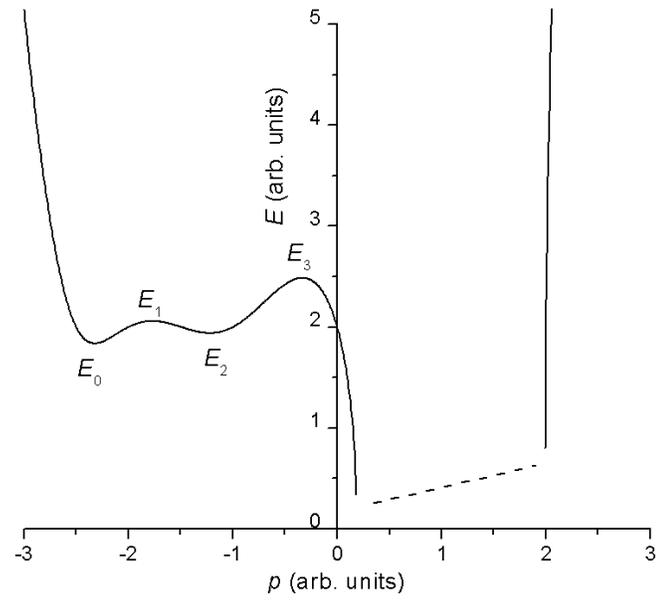

**FIG. 2.** An example of a sub-band with an arbitrary number of local extrema.

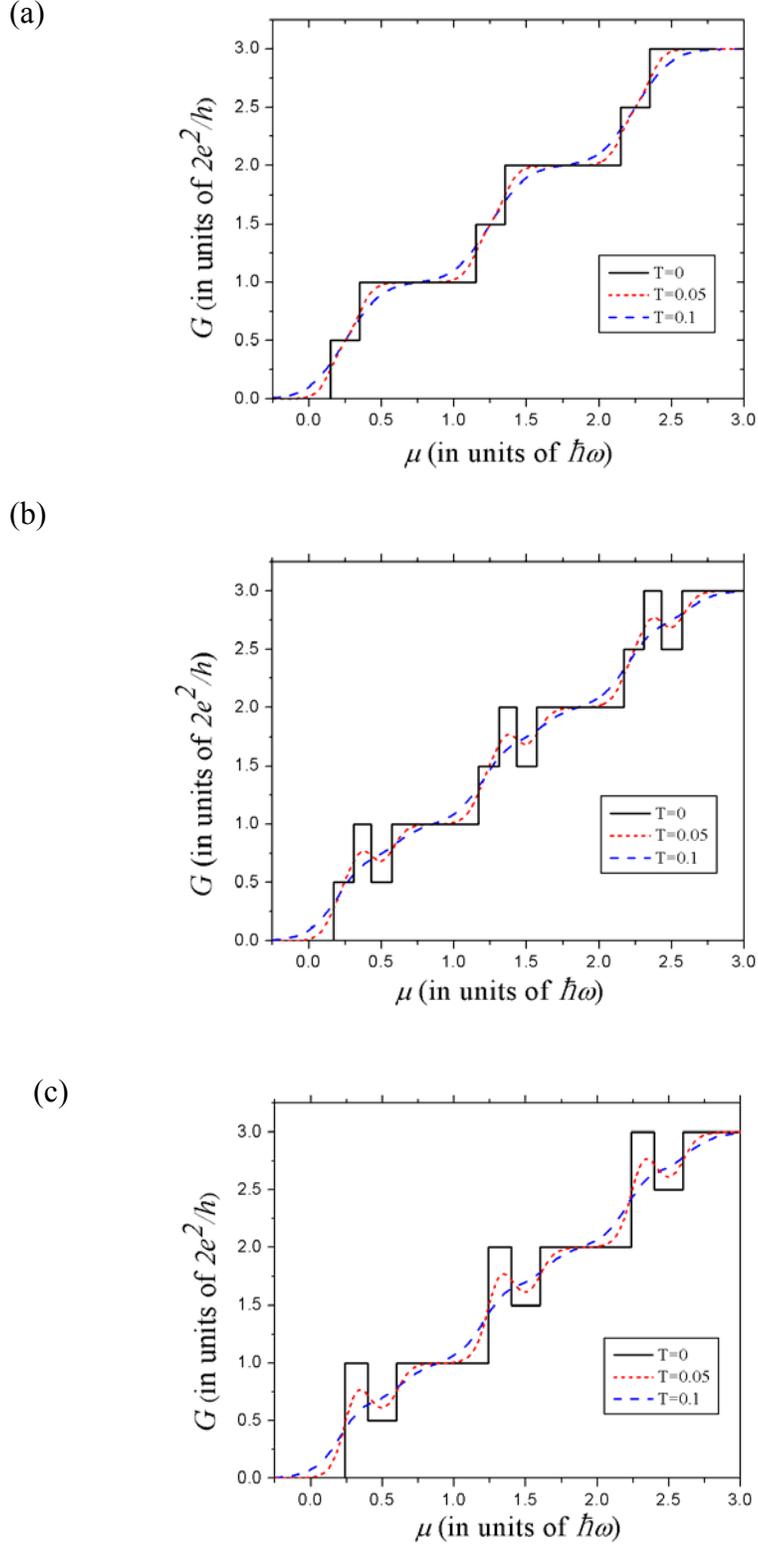

**FIG. 3.** Conductance as a function of the chemical potential at different temperatures, T (in units of $\hbar\omega/k_B$), for different directions of the magnetic field: (a) $\theta = \pi/2$, (b) $\theta = \pi/4$, and (c) $\theta = 0$.